\documentclass[letter,twocolumn]{IEEEtran}
\usepackage{amsfonts}
\usepackage{times}
\usepackage{graphicx}
\usepackage{latexsym}
\usepackage{dsfont}
\usepackage{amssymb}
\usepackage{amsmath}
\usepackage{cite}
\usepackage{verbatim}
\usepackage{subfigure}

\newcommand{\figref}[1]{{Fig.}~\ref{#1}}

\def\bb0{{\mathbb{0}}}

\def\ba{{\mathbf{a}}}
\def\bb{{\mathbf{b}}}

\def\bff{{\mathbf{f}}}

\def\bh{{\mathbf{h}}}

\def\bo{{\mathbf{o}}}

\def\b0{{\mathbf{0}}}

\def\bO{{\mathbf{O}}}

\def\bX{{\mathbf{X}}}

\def\sf0{{\mathsf{0}}}

\usepackage{epstopdf}
\usepackage{enumerate}
\usepackage{algorithm}
\usepackage{amsmath}
\usepackage{float}
\usepackage{color}
\usepackage{makeidx}
\usepackage{bm}
\usepackage{cleveref}
\usepackage{url}
\usepackage{steinmetz}
\usepackage{varwidth}
\usepackage{soul}
\graphicspath{{figs/}}
\usepackage{lipsum}
\usepackage{multicol}
\usepackage{multirow}
\usepackage{booktabs}

\DeclareMathOperator*{\argmax}{arg\,max}
\DeclareMathOperator*{\argmin}{arg\,min}

\usepackage{bbm}

\begin{document}

\title{Enabling ISAC in Real World: Beam-Based User Identification with Machine Learning}

\author{Umut~Demirhan~and~Ahmed~Alkhateeb \thanks{The authors are with the School of Electrical, Computer and Energy Engineering, Arizona State University, Tempe, AZ, 85281 USA (Email: udemirhan, alkhateeb@asu.edu). This work was supported by the National Science Foundation (NSF) under Grant No. 2048021.}}

\maketitle

\begin{abstract}
	Leveraging perception from radar data can assist multiple communication tasks, especially in highly-mobile and large-scale MIMO systems. One particular challenge, however, is how to distinguish the communication user (object) from the other mobile objects in the sensing scene. This paper formulates this \textit{user identification} problem and develops two solutions, a baseline model-based solution that maps the objects angles from the radar scene to communication beams and a scalable deep learning solution that is agnostic to the number of candidate objects. Using the DeepSense 6G dataset, which have real-world measurements, the developed deep learning approach achieves more than $93.4\%$ communication user identification accuracy, highlighting a promising path for enabling integrated radar-communication  applications in the real world. 
\end{abstract}


\section{Introduction}
Employing large antenna arrays in millimeter wave (mmWave) communication systems is essential to provide sufficient beamforming gains and receive power. Minimizing the beam training overhead, however, is a challenging task especially in highly-mobile applications. Towards addressing this challenge, integrated sensing and communications (ISAC) can be a keystone, where wireless communications are aided by radar sensing information. This information can be utilized to build perception about the environment and the target objects that affect the communication channels. One main challenge in this framework, however, is that the objects determined in both the sensing  and communication domains need to be matched to facilitate accurate sensing aid to communication in scenarios with multiple mobile objects in the sensing scenes. We term this problem \textit{user identification}. In this paper, we investigate how to solve the user identification problem in the radar sensing scenes via the use of selected beam information and evaluate the feasibility using a real-world demonstration.

Various types of sensing information, e.g., camera images \cite{Alrabeiah2020a}, and radar \cite{ali2019millimeter,  demirhan2022radarbeam, demirhan2022radarblockage, liu2020radar}, have been considered for aiding communication objectives. In \cite{Alrabeiah2020a}, images captured by an RGB camera attached to the basestation are considered for the beam prediction and blockage prediction tasks. Similarly, the measurements from an out-of-band radar have been studied for beam prediction \cite{ali2019millimeter, demirhan2022radarbeam} and blockage prediction tasks \cite{demirhan2022radarblockage}. To realize radar-aided systems in the real world, however, there are many major questions to be answered.  To this end, although \cite{demirhan2022radarbeam, demirhan2022radarblockage} included an evaluation in a real-world setup, they were mainly limited to the single-target scenarios, and the identification of the user has not been considered. In the joint sensing and communication literature, where the sensing and communication antennas are common, there has been some work for the user identification \cite{liu2020radar, aydogdu2020distributed, wang2022multi}, where the users are identified based on Euclidean distance \cite{liu2020radar} and Kullback-Leibler divergence \cite{wang2022multi} of the radar/communication estimated positions, and GPS positions \cite{aydogdu2020distributed}. These works, however, did not (i) consider an off-the-shelf external radar, (ii) evaluate the performance in a real-world system, (iii) utilize machine learning, which may be a key for ISAC systems \cite{demirhan2022integrated}, and (iv) use the beam index of the communication system for identification, which is readily available in the deployments.

Therefore, in this work, we first define the user identification problem in radar-aided communication systems that use out-of-band frequency modulated continuous wave (FMCW) radars. Specifically, using the radar-generated measurements of the detected objects and the beam indices selected to serve the communication user, we develop a robust and scalable deep neural network (DNN) solution for this problem. For the evaluation, we build a real-world dataset as a part of the Deepsense 6G framework \cite{DeepSense} and compare our approach with baseline solutions on this dataset. In our evaluations, the proposed DNN solution offers $93.4\%$ accuracy with over $\%20$ gain over the baselines and shows promising potential for real-world radar-aided communication systems.

\section{System Model}

\begin{figure}[!t]
	\centering
	\includegraphics[width=1\columnwidth]{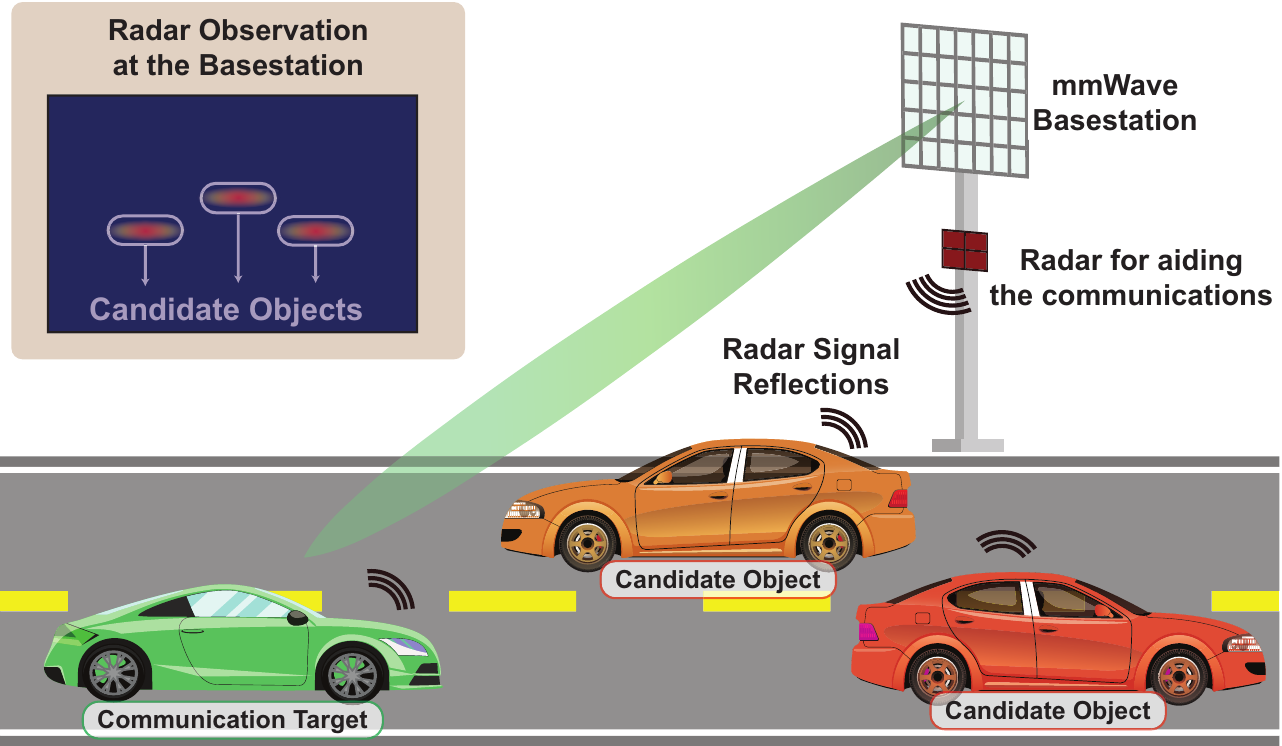}
	\caption{The system model, where a radar-equipped mmWave basestation communicates with a user, while radar data to aid the communications is collected from all the available targets in the environment.}
	\label{fig:systemmodel}
\end{figure}

For the system model, as illustrated in \figref{fig:systemmodel}, we consider a single mobile user served by a basestation, along with multiple candidate targets. The mobile user carries a single antenna mmWave receiver. Meanwhile, the basestation is equipped with (i) a mmWave antenna array that is used to communicate with the mobile user and (ii) an off-the-shelf FMCW radar that is leveraged to aid the mmWave communication functions.

\subsection{Communication Model}
For communications, we consider a MISO channel with $N$ element antenna array at the basestation. For simplicity, in the following, we assume that the transmitter antennas form a uniform linear array (ULA). Nevertheless, this assumption can easily be relaxed by extending to any other antenna formation. Adopting a geometric channel model channel with $P$ paths, we can write \cite{demirhan2022radarbeam}
\begin{equation}
\bh = \sum_{p=1}^P \alpha_p \ba(\theta_p),
\end{equation}
where $\alpha_p$ and $\theta_p$ denote the complex coefficient and azimuth angles of the $p$-th path. The function $\ba(\theta)$ is defined as the array response vector of the basestation antenna array in the direction of $\theta$.  At the downlink, the basestation applies a beamforming vector $\bff \in \mathbb{C}^{N}$ to the information $s$ and transmits this signal over the channel. Hence, the signal received by the user can be written as
\begin{equation}
	y = \sqrt{\rho} \bh^H \bff s + n,
\end{equation}
with $\rho$ being the transmitter power gain of the basestation and $n \sim \mathcal{N}(0, \sigma^2)$ being the additive white Gaussian noise. The beamforming vector $\bff$ is selected from a codebook $\bm{\mathcal{F}}$ of $B$ beams. The $b$-th beamforming vector of the codebook is denoted by $\bff_b$. With this model, the index of the optimal beam can be obtained by the beamforming gain maximization problem given as
\begin{equation}
	b^\star = \argmax_{b \in \{1, \ldots, B\}} \lvert \bh^H \bff_b \rvert^2,
\end{equation}
where the optimal solution can be obtained by an exhaustive search over the beams of the codebook.

\subsection{Radar Model}
As described, our system model adopts an FMCW radar at the basestation. This radar aims to provide measurements to aid the communication system. Specifically, the radar transmits a sequence of linear up-chirps, given as \cite{li2021signal}
\begin{equation}
s^{\textrm{tx}}_{\textrm{chirp}} (t) = \sin(2\pi [f_c t + \frac{\mu}{2} t^2]),
\end{equation}
for $t \in [0, T_c]$, and $0$ otherwise. The chirp signal starts from the frequency $f_c$ and goes up to $f_c + \mu t$ with a constant slope $\mu=B/T_c$ within the duration of the signal. $B$ and $T_c$ represent the bandwidth and duration of a chirp signal.

The transmission of a sequence of chirps comprises a radar frame. Specifically, a sequence of $M_c$ chirps, with $T_w$ waiting time between two consecutive chirps, is transmitted in each radar frame. With the transmission of a radar frame, the signal is reflected from the objects in the environment and received back at the receiver. The FMCW radar receiver mixes the current transmit signal with the received signal. The resulting signal is called an intermediate frequency (IF) signal. If there is a single object in the environment at a $d$ distance, the IF signal from a single chirp can be written as
\begin{equation}
s_\textrm{chirp}^\textrm{rx}(t) = \sqrt{\mathcal{E}_t \mathcal{E}_r} \exp\left(j2\pi \left[\mu \tau_{rt} t + f_c \tau_{rt} - \frac{\mu}{2} \tau_{rt}^2\right]\right),
\end{equation}
where $\mathcal{E}_r$ is the channel gain comprised of the gain due to the reflection/scattering and path-loss. $\tau_{rt} = 2d/c$ represents the round-trip delay of the reflected signal with $c$ being the speed of light. The IF signal is sampled by an analog-to-digital converter (ADC) at the sampling rate, $f_s$, producing $M_s$ samples per chirp per RF chain. Given the $M_c$ chirps per frame and assuming a radar with $M_a$ receiver antennas (with an RF chain for each antenna), a radar measurement produces $M_a M_c M_s$ ADC samples. We denote the ADC samples of each radar measurement (of a frame) by $\bX^r \in \mathbb{C}^{M_a \times M_c \times M_s}$.

Further, a classical object detection method (e.g., \cite{li2021signal}) is applied to the radar measurements, $\bX^r$, to detect the objects as follows. (i) The range FFT, clutter cleaning, angle FFT, and Doppler FFT are applied to obtain a radar cube. Note that these FFTs are through the three dimensions of the radar measurements. (ii) Using the radar cube, a CFAR method detects the points with high power. (iii) These detected points are clustered with DBSCAN to determine the groups of detected points representing the candidate objects. (iv) The range $o^r_k$, angle $o^a_k$, and Doppler velocity $o^v_k$ of each cluster $k$ are estimated as the average of the cluster's points. Specifically for each object, we have the properties $\bo_k = [o^r_k, o^a_k, o^v_k]$. After this operation, we now have $K$ objects, which include the communication target. 

\section{Problem Definition: User Identification with Beam Information}
In this work, we aim to determine the communication target within the radar measurement, which is an imperative task for using radar sensing for communications. An important observation here is that this task requires a piece of communication information for the mapping between the radar and communication domains. For this purpose, we use the optimal beam index of the communication target, which may be obtained by initial access/channel estimation (mainly at sub-6 GHz) or beam sweeping (at higher frequencies). Then, given the beam index, we aim to find the communication user within the radar targets.

To formalize the problem, we first denote the number of samples by $T$ and introduce the sub-index $t$ to our notation. We then assume that there are $K_t$ candidate objects in the radar measurement $\bX^r_t$, which includes the communication target. Let us denote the index of this communication target by $k_t^c \in \{1, \ldots, K_t\}$. Recall that for each sample $t \in \{1, \ldots, T\}$, we have the radar measurements of the objects, $\bO_t = [\bo_{t,1}, \ldots, \bo_{t,K_t}]$, and the optimal beam index of the communication target, $b^\star_t$. Given this information, our purpose is to determine the index of the communication target, $k^c_t$. 

Mathematically, let us assume that there exists a function of parameters $\boldsymbol{\Theta}$ that maps the radar measurement and the optimal communication beam index to the radar association information of the target object, i.e., 
\begin{equation} \label{eq:function}
	f_{\boldsymbol{\Theta}}(\bO_{t}, b^\star_t) = k^c_t. 
\end{equation}
Then, our aim becomes to approximate this function and parameters with minimal error, which can be formulated as
\begin{equation} \label{eq:objective_main}
	\hat{f}_{\hat{\bm{\Theta}}} = \argmin_{f, \bm{\Theta}} \frac{1}{T} \sum_{t=1}^T \mathcal{L}\big(f(\bO_t, b^{\star}_t), k^c_t\big),
 \end{equation}
for a given loss function $\mathcal{L}$. With this objective in mind, different approaches could be developed. For instance, a machine learning model can be trained using \eqref{eq:objective_main} directly. As an alternative, classical signal processing-based solutions may be proposed via the design of a fixed $f_{\boldsymbol{\Theta}}$ function.

A crucial step in this formulation is to determine the index of the communication target. To achieve this, one can mark the communication object by inspecting the radar data with the aid of other potential data modalities (e.g., camera images), as applied in this paper. For actual deployments, more scalable and less error-prone options can be developed. An example could be by transmitting specifically designed signals at the radar frequency from the target object. These signals can be detected at the radar receiver in addition to the radar measurements, and the target object can be marked. Such approaches, however, may also result in interference, so the details of their design should be carefully evaluated. Next, we propose our solutions in the following section.

\section{Proposed Solutions}
For the user identification problem, we develop two approaches: (i) A model-based baseline solution that maps the angle of the object in the radar to the angle of the communication beam, and (ii) a scalable machine learning solution that is agnostic to the number of candidate objects. Next, we present these two developed solutions. 

\textbf{Baseline Solution:}
An important observation on the communication beamforming with the directional beams (e.g., DFT codebook) is that they point to the quantized angle of the communication object. Considering a line-of-sight channel, this information can be directly mapped to the radar angle. In this mapping, however, the angular misalignment of the radar and communication antennas may cause a significant error. Thus, we design a radar-to-communication angle mapping solution by considering the misalignment. In this solution, the angle corresponding to the communication beam index is mapped to the radar angle of the communication target, and the candidate object with the closest angle is selected as the communication target.

To formalize the approach, we denote $\psi^r_{t, k}=o^a_{t,k}$ and $\psi^c_{t, k}$ as the radar and communication angles of the $k$-th candidate object in sample $t$. We also denote the pointing angles of the communication beams by $\phi_b, \forall b \in \{1, \ldots, B\}$. With the proposed model and notation, the communication angle of the target object $\psi^c_{t, k_t^c}$ will have the minimum angle difference to the optimal communication beam angle, as it provides the maximum SNR. Therefore, we take $\psi^c_{t, k_t^c} \approx \phi_b$. This approximation particularly holds with a large number of beams, as they are narrower.

To convert the communication angles of the candidate objects to the radar angles, we adopt a simple model based on the angle offset (misalignment) between the radar and communication antennas. If $\psi_0$ denotes this angle offset, the transformation between the communication and radar angles of an object can be written as
\begin{equation} \label{eq:angleoffset}
	\psi^r_{t, k} = \psi_0 + \psi^c_{t, k}, \ \forall k \in \{1, \ldots, K_t\}.
\end{equation}
To estimate the offset, $\psi_0$, we utilize mean squared error (MSE) over the data samples of the communication targets by $\hat{\psi}_0 = \frac{1}{T} \sum_{t=1}^T \lvert \psi^c_{t, k_t^c} - \psi^r_{t, k_t^c}\rvert^2$. Finally, to determine the index of the communication target, we select the candidate object that has the smallest distance to the radar angle of the target object, i.e.,
\begin{equation}
	\hat{k}^c_t = \argmin_{k \in \{1, \ldots, K_t\}} \lvert \psi^r_{t, k_t^c} - \psi^r_{t, k} \rvert.
\end{equation}


We want to note that the approach developed in this section assumes a line-of-sight channel and considers no other error than the misalignment of the radar and communication antennas. This approach could be expected to perform well in simulation-based models. In real systems, however, various imperfections (e.g., beam gain response of the actual hardware), in addition to the potential angle detection errors of the radar targets, may cause the approach to perform poorly. In the following, we aim to develop a robust solution that can adapt to these potential errors.

\begin{figure}[!t]
	\centering
	\includegraphics[width=1\columnwidth]{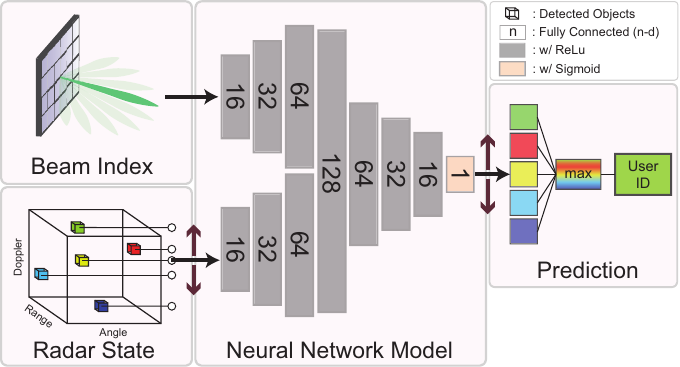}
	\caption{The proposed DNN solution. Along with the communication beam index, the radar estimated state of each user is respectively fed to the DNN. The resulting outputs are collected together to select the communication target.}
	\label{fig:dnn-model}
\end{figure}

\begin{figure*}[!t]
	\centering
	\includegraphics[width=1.95\columnwidth]{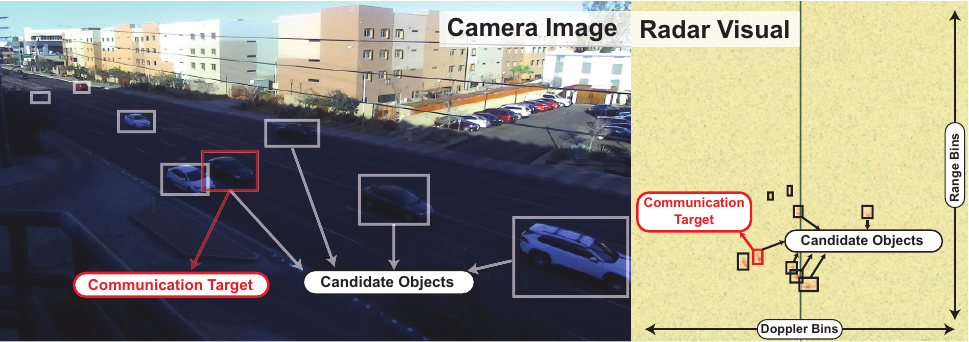}
	\caption{A sample from the dataset is shown with the corresponding camera image and range-Doppler map. In this sample, the beam with the index $28$ corresponds to the optimal beam. The targets are determined using this visual, based on a sequence of samples by tracking through time.}
	\label{fig:dataset_sample}
\end{figure*}

\textbf{Deep Learning Solution:} As mentioned, different types of imperfections, including object detection errors, may be present in real-world systems. To develop a robust solution, in this section, we propose a deep neural network model. This way, the imperfections can be learned and accommodated with the DNN models. Further, the additional information that has not been utilized in the baseline approach, i.e., the range and Doppler velocity of the objects, can be beneficial. Although the DNNs adapt to the imperfections by taking advantage of the available data, their design for this problem is not a straightforward task.

There are two important challenges with the design: (i) The number of candidate objects, $K_t$, is a sample-dependent parameter and can have various values at different instances. For this, one can design a model taking a predetermined maximum number of given objects as the input and append $0$s to have an input of the predetermined size. Such an approach, however, could be a burden on the model's complexity. (ii) A desirable property of such a solution is permutation invariance. A solution that takes the information of all objects at once as an input may also utilize the ordering information of the objects, which is not desirable. Although this may be resolved by generating more samples by mixing the order of the objects, it possibly requires significantly more training for a large number of objects. \textit{Overcoming both challenges, we propose our solution, where the radar information of each candidate object and the optimal radar communication index are respectively fed to the same neural network}. This network then returns the likelihood of the candidate object being the communication target. These soft predictions (likelihood ratios) are collected together for each candidate object, and the candidate object with the maximum probability is returned as the target object. We illustrate our approach in \figref{fig:dnn-model}.

In this solution, the neural network takes four values as the input, i.e., the optimal beam index $b^\star_t$, and range, angle, and Doppler velocity, $\bo_{k, t}$. As shown in \figref{fig:dnn-model}, in the architecture, we adopt a set of $7$ layers. The first three layers are composed of two separate lines of fully connected layers, each three expanding the input of radar information and beam index, respectively, with the aim of extracting potential features for combining. The fourth layer takes the outputs of the lines of three layers and combines the extracted features of the radar and communication. The following layers reduce the dimension and return the prediction. For the training of the model, we use
\begin{equation}
\argmin_\mathbf{\Theta} \frac{1}{T} \sum_{t=1}^T \sum_{k=1}^{K_t} \mathcal{L}(g_\Theta(\bo_{t,k}, b^\star_t), y_{t,k})
\end{equation}
where $g_\mathbf{\Theta}$ represent the neural network function and constructs the function defined in \eqref{eq:function} by $f_\Theta(\bO_{t}, b_t^\star) = \argmax_k g_\Theta(\bo_{t,k}, b^\star_t)$. The variable $y_{t,k}$ is the indicator of the candidate object being the communication target, i.e., $y_{t,k}=1$ if $k=k^c_t$ and $0$ otherwise. From the objective, it can be seen that different candidate objects are treated as different samples. As the loss function, we utilize the MSE loss function. Next, we present our real-world dataset for the evaluation.

\section{Dataset}
For a realistic evaluation of the proposed user identification solutions, we built a real-world dataset. In collecting this dataset, we used a hardware testbed with co-existing radar and wireless mmWave equipment, following the DeepSense dataset structure \cite{DeepSense}. Then, by processing the collected measurements/raw dataset, we built our development dataset for user identification in radar signatures. In this section, we describe our testbed and development dataset.

\textbf{Testbed:} We adopt Testbed 5 of the DeepSense 6G dataset \cite{DeepSense} for the data collection, similar to \cite{demirhan2022radarbeam, demirhan2022radarblockage}. Testbed 5 comprises two units: (i) Unit 1, a fixed receiver acting as a basestation, and (ii) Unit 2, a mobile transmitter representing the target object. The rest of the details can be found in \cite{demirhan2022radarbeam}. Differently from this work, the radar chirp parameters are selected based on long-range, providing a maximum range of $249$m and velocity of $82$ km/s. The bandwidth of the utilized chirp frame covers $B=310$ MHz bandwidth with a chirp slope of $\mu=10$ MHZ/us over $L=250$ chirps/frame and $S=512$ samples/chirp.

\textbf{Development Dataset:}
We construct Scenario $35$ of the DeepSense dataset \cite{DeepSense}. In this scenario, a base station is placed on the second floor of a parking spot, directed towards a road with dense traffic. With this placement, we aimed to create a realistic and challenging data collection scenario. During the collection, a car with the transmitter is driven through the road in both directions. The received power via each beamforming vector and radar measurements are saved continuously at the rate of $9$ samples/s, to be processed later. For the construction of the dataset, the beam providing the most power and the corresponding power level is saved as the optimal beamforming vector.

For the processing, we determined the set of samples where the communication target is in the scene utilizing the optimal beam's power. Specifically, long sequences containing sufficient receive power and a roughly linear optimal beam index pattern are included in the dataset. This intermediate dataset without labels contained $3045$ samples. For the labels of the target objects, we applied a classical object detection solution. We then visually inspected the radar maps along with the RGB images. Through the evaluation of the samples over time, the objects that can be certainly identified as the target objects are marked. The final dataset contained $2158$ samples. To prevent over-fitting to specific sequences (unique passes of the car), we split the data $80/20\%$ training and test sets based on their sequences through the field of view. Next, we evaluate our solutions using this dataset.

\section{Results}

In this section, we evaluate the proposed solutions. In addition, we generalize the baseline solution and include three additional approaches: (i) Angle-based linear regression, where we use $\psi^r_{t, k} = \psi_0 + \alpha \psi^c_{t, k}$ for a parameter $\alpha$ instead of \eqref{eq:angleoffset}, (ii) linear regression for the whole radar state (range, angle, and Doppler velocity), and (iii) beam-radar angle lookup table, where each beam is mapped to the average radar angle of training samples, and the candidate object with the closest angle to the angle of the beam is selected as the communication target. For these, we estimate the parameters by utilizing the training set.  For the training of the DNN solution, we used the ADAM algorithm with a learning rate of $0.001$. The network is trained for $100$ epochs with a batch size of $32$. The final weights after all epochs are adopted for the evaluation of the solution. We now present our results.

\begin{figure}[!t]
	\centering
	\includegraphics[width=1\columnwidth]{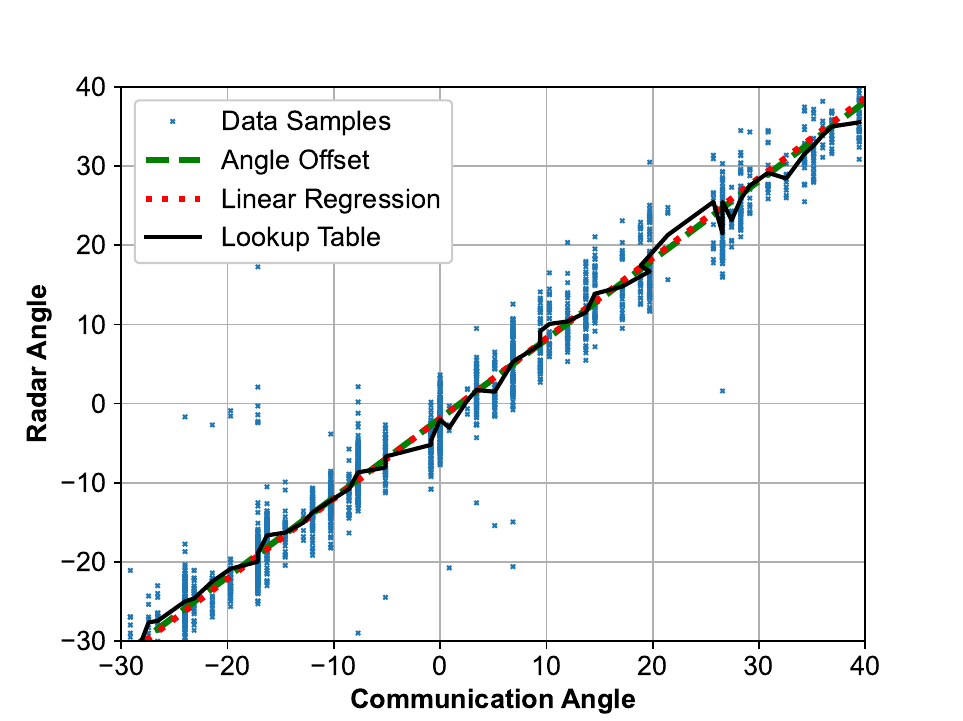}
	\caption{The visualization of the data samples and the baseline solutions. Most of the samples follow a linear pattern with a relatively large variance.}
	\label{fig:dataset_visualization}
\end{figure}

First, in \figref{fig:dataset_visualization}, we illustrate the radar and beam angles of the samples in the dataset along with the baseline approaches. The data samples, on average, show a linear pattern and support the linear model design. However, there are variations in the samples, which may be due to the detection and other practical errors. This may cause performance degradation for the model. To that end, the DNN solution is expected to perform better by adjusting to the data.

In Table \ref{tab:performance}, we show the accuracy of detecting the target object among the candidates. The angle offset-based baseline solution delivers a low performance. On the other hand, linear regression significantly improves the result with $\alpha=0.9877$, where a slight mismatch between the angle scaling is observed. The lookup table further enhances the performance, however, they cannot provide performance close to the DNN solution at $93.4\%$. This indicates the significant advantage of capturing the non-linear imperfections and errors with the proposed DNN solution. Furthermore, user identification is tightly coupled with the potential radar-aided communication solutions, and its errors can be amplified within the whole solution, rendering the approach infeasible. To that end, the high accuracy of the proposed deep learning approach enables radar-aided communication solutions with high accuracy.

\begin{table}[]
\caption{User Identification Performance }
\centering
\setlength{\tabcolsep}{12pt}
\renewcommand{\arraystretch}{1.5}
\begin{tabular}{lc}
\textit{\textbf{Solution}} & \textit{\textbf{Accuracy}} \\ \hline
Angle Offset (Baseline)     & 0.5354  \\ \hline
Linear Regression (Angle)          & 0.6850   \\ \hline
Linear Regression  (3D)        & 0.6955   \\ \hline
Lookup Table               & 0.7244                     \\ \hline
Deep Learning              & 0.9343                    \\

\end{tabular}
\label{tab:performance}
\end{table}

\section{Conclusion}
Radar-aided communications can be essential in advancing the performance of communication systems. One particular problem in aiding the communication with the radar is determining the relevant radar data to the communication target. In this paper, we formulated this problem and developed alternative solutions, including a DNN-based scalable approach. The DNN solution achieved $93.4\%$ accuracy by significantly outperforming the baseline solutions and presented a promising result for enabling radar-aided communication systems.

\bibliographystyle{IEEEtran}

\end{document}